\documentclass{elsart}

\usepackage{psfig}

\usepackage{natbib}

\def\ls{{_<\atop^{\sim}}}
\def\gs{{_>\atop^{\sim}}}

\begin{document}

\runauthor{F. Nicastro}


\begin{frontmatter}

\title{A New Model for the Width of the Broad Emission Lines in AGN: 
The Role of the Accretion Rate}

\author[PSU]{F. Nicastro and M. Elvis}
\address[PSU]{Harvard-Smithsonian Center for Astrophysics, 
60 Garden Street, Cambridge, MA, 02138, USA}

\begin{abstract}
We present a model which relates the Broad Emission Line 
Regions of AGN to the accretion mechanism. Based on the Keplerian 
assumption for the high velocity clouds of the gas responsible for 
the production of the Broad Emission Lines, we identify 
that gas with an outflowing disk/coronal wind forms a stabilized 
Shakura-Sunyaev accretion disk. 
We show that in the framework of this picture the observed range of 
H$\beta$ FWHM from Broad Line to Narrow Line type 1 AGN, is well 
reproduced as a function of the accretion rate. 
\end{abstract}


\end{frontmatter}


\section{Introduction}
The main facts about Broad Emission Line Regions (BELRs) of AGNs 
can be summarized as follow: 
{\bf (a) Ubiquity:} Broad Emission Lines (BELs) are probably 
ubiquitous in AGN, being unobserved only when a good case exists 
for their obscuration by dust (type 2 AGN) or their being 
swamped by beamed continuum (Blazars). 
This suggests that the presence of Broad 
Emission Line Clouds (BELCs) in the AGN environment is closely 
related to the mechanisms which are responsible for the quasar 
activity. {\bf (b) Photoionization:}
reverberation studies of the BELs in many Seyfert 1 galaxies 
(e.g. Wandel, Peterson \& Malkan, 1999: hereinafter WPM99) 
indicate that the gas of the BELCs is photoionized, and that 
its physical state, from one object to another, covers a rather 
narrow range of parameter space (in column density, electron 
density, ionization parameter). {\bf (c) Virialization:}
within a single object the BELCs are stratified, with the highest 
ionization lines being also the broadest (Krolik et al., 
1991; Peterson et al., 1999). 
This accords with the photoionization hypothesis 
and with the idea that the BELs are broadened by their orbital 
Keplerian motion around the central source (Peterson \& Wandel, 1999). 
{\bf (d) Wide Range of Velocity:}
BELs do not at all have the same dynamical 
properties in all objects. A broad distribution of line widths 
from $\sim 1,000$ km s$^{-1}$ (in Narrow Line Seyfert 1 galaxies, 
NLSy1) to $\sim 20,000$ km s$^{-1}$ (in the broadest broad line type 
1 AGN) is present. 

Here we briefly present a model (Nicastro, 2000, hereinafter N00), 
based on the Keplerian assumption and which links the existence of 
the BELRs with the mechanism responsible for the quasar activity: 
accretion onto a massive black hole. Formulae, numerical details 
and definitions can be found in N00. 
The model explains naturally (a) the ubiquity of BELs 
in AGNs, and (b) the wide (but limited) range of velocities observed. 
Stratification in a single object can also be, at least partly, 
accounted for by this model. 

\vspace{-.5cm}
\section{The Model}
We adopt the scenario described by Witt, Czerny and Zycki (1997, 
hereinafter WCZ97), and its recent revision (A. Rozanska and B. Czerny 
2000, in preparation). These authors studied a radially co-accreting 
disk/corona system and demonstrated that a transonic vertical outflow 
from the disk develops where the fraction of total energy dissipated 
in the corona is close to the maximum, and for accretion rates higher 
than a minimum value, below which evaporation inhibits the formation 
of the wind. 
In the framework of this model, and consistent with the empirical 
quasar structure presented by Elvis (2000) at this conference, 
we propose that a vertical disk/coronal wind, is the origin of the 
BELCs and that the widths of the BELs are the Keplerian velocities of 
the accretion disk at the radius where this wind arises (N00). 
In our model the disk wind forms for external accretion rates 
higher than a minimum value $\dot{m}_{min}(m)$ (in critical units; see N00 for 
details) below which a standard Shakura-Sunyaev (1973, hereinafter SS73) 
disk (i.e. SS-disk) is stable and extends down to 
the last stable orbit. 
The model explains the observed range of FWHM in the BELs of AGN as 
a function of a single physical quantity connected with the AGN activity: 
the accretion rate. 

{\bf Width of the Broad Emission Lines:} 

We adopt an average radius $r_{wind}$ for the transonic 
outflow (and so for the BELRs), obtained by weighting the radial distance 
by $(1-\beta)$, between $r_{tran}$ and $r_{max}$ (see equations 1-3 in 
N00 for appropriate definitions). 
In Figure 1 we show the relationship between the accretion rate (in the 
range $\dot{m}_{min}(m) - 10$) and the expected FWHM$(r_{wind})$ 
(solid, thick curves) and FWHM($r_{max}$) (dashed, thin curves), for 
$m = 10^6, 10^7, 10^8, 10^9$ based on the Keplerian assumption. 
These curves are independent of the mass of the 
central black hole, and so they overlap in the diagram of 
Figure 1. 
However, for each curve, the maximum achievable FWHM depends on the 
mass, via the limit imposed by the minimum external accretion 
rate $\dot{m}_{min}(m)$ needed for an unstable SS-disk to exist. 
At the bottom of the plot the four horizontal lines indicate these 
values of $\dot{m}_{min}(m)$. 
The parameter space between the dashed and solid curves of Figure 1 
gives a range of $\sim 1.4$ in FWHM (i.e. $\sim 2$ 
in radius) for a given accretion rate, so allowing, at least partly, 
for a stratification of the BELCs in a single object. 
For accretion rates $\dot{m} < 0.2$ (sub-Eddington regime) the 
predicted FWHM are quite broad ($\gs 4,000$ km s$^{-1}$), and similar 
to those typically observed in broad line type 1 AGN (Figure 1). 
For $\dot{m} = 0.2 - 3$ (Eddington to moderately super-Eddington) the 
corresponding FWHM span the interval $\sim 1,000 - 4,000$ km s$^{-1}$, which 
contains the value of FWHM$ = 2,000$ km s$^{-1}$ used to separate 
normal type 1 AGN and NLSy1. 
Hence our model predicts that narrow line type 1 AGN accrete at 
higher accretion rates compared to broad line objects, as suggested by  
Pounds et al. (1995). 
However, the mass of the central black hole in NLSy1 does 
not need to be smaller than that of broad line type 1 AGN. 
No instability is possible for $\dot{m} \ls \dot{m}_{min}(m)$ thus 
providing an upper limit of $\sim 15000-20000$ km s$^{-1}$ for the 
BEL FWHM. 
\begin{figure}[htb]
\centerline{\psfig{figure=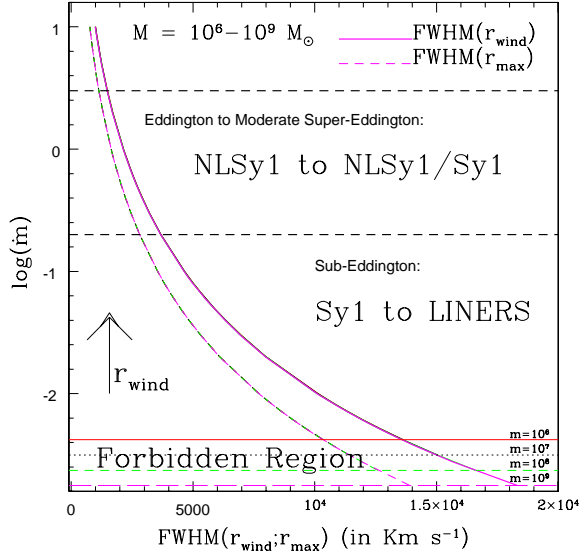,height=3.5truein,width=3.5truein,angle=0}}
\caption{Predicted accretion rate vs FWHM$(r_{wind})$, FWHM($r_{max}$), 
for $m = 10^6, 10^7, 10^8, 10^9$, and $\dot{m}$ in the range 
$\dot{m}_{min}(m) - 10$.}
\end{figure}

{\bf Relative Radiative Efficiency:} 

Using the predictions of our model, and the recent measurements of 
distances and FWHM reported by Wandel, Peterson and Malkan (1999) 
for a sample of 19 type 1 AGNs, we calculate the ratio 
$\xi$ between the measured (WPM99) dimensionless luminosity 
$\ell = (L_{ion}/L_{Edd})$ and the predicted accretion rate. 
$\xi$ is then a measure of the relative efficiency (compared to the 
maximum radiative efficiency $\eta$: $\xi = \epsilon / \eta$, where 
$\epsilon = L_{ion}/\dot{M}c^2$) by which the accretion power is 
converted into ionizing luminosity.
We found that $\xi$ is correlated with the measured mass of the central 
black hole (see Figure 3 in N00). 
This observational result is not an obvious consequence of our model 
and needs further study. 
This correlation may explain why we do not see very-low mass 
AGN ($\ls 10^4$ M$_{\odot}$): the efficiency in converting accretion 
power into luminosity may be too low to observe such objects. 
%
%

\section{Conclusions}
We presented a simple model which tightly links the existence of the 
BELCs in type 1 AGN to the accretion mechanism (N00). 
We identify the BELR with a vertically outflowing wind of ionized 
matter which forms at a critical radius in the accretion disk, 
consistent with the model for quasar structure presented by Elvis 
(2000) at this conference. 

Our main findings are: (a) the entire observed range of velocities 
(FWHM) in type 1 AGN is naturally reproduced in our model by  
allowing the accretion rate to vary from its minimum permitted 
value to super-Eddington rates; (b) for accretion rates close to 
the Eddington value, the expected FWHM are of the order of those 
observed in NLSy1. Lower accretion rates give instead FWHM typical  
of broad line type 1 AGN. (c) We find an empirical relationship which 
suggests that the radiative efficiency of higher mass black holes is 
greater than that for lower mass black holes. 




\end{document}